\documentclass[%
 reprint,
 amsmath,amssymb,
 aps,
prstab,
]{revtex4-1}
\usepackage{makecell}
\usepackage{graphicx}
\usepackage{dcolumn}
\usepackage{epstopdf} 
\usepackage{bm}

\begin{document}
\preprint{APS/123-QED}

\title{Horizontal Emittance Reduction on a Synchrotron Radiation Light Source with a Robinson Wiggler}
\author{H. Abualrob}
\email{hadil.abualrob@najah.edu}
 \altaffiliation[Also at]{ An-Najah National University, Nablus, Palestine}
\author{P. Brunelle, M.-E. Couprie, M. Labat, A. Nadji, L. S. Nadolski}%
\author{O. Marcouille}
\affiliation{Synchrotron SOLEIL, L'Orme des Merisiers, Gif-sur-Yvette, France}

\date{\today}
\begin{abstract}
The performance of synchrotron light facilities are strongly influenced by the photon beam brightness, that can be further increased by reducing the beam emittance. A Robinson Wiggler can be installed in a non-zero dispersion straight section to reduce the horizontal beam emittance. It is composed of an array of magnets of alternated polarities, whose both magnetic field and gradient are of opposite signs. It provides a compact solution to reduce by 50$\%$ the horizontal emittance. However, it increases the energy spread by 40$\%$. The concept of the Robinson Wiggler (RW) is described here, the first experimental observation of the Robinson effect in a synchrotron light source on the transverse and longitudinal beam properties by the means of four existing undulators is presented and the impact on the photon flux density is studied.

\end{abstract}

\pacs{Valid PACS appear here}

\maketitle 
 
\begin{center}
\textbf{INTRODUCTION}\\
\end{center}

High brightness light sources have allowed for the development or the improvement of new techniques \cite{lengeler2001coherence, huang2013brightness}, e.g. coherent X-ray diffraction imaging \cite{Miao:1999aa} and holography \cite{kondratenko1977use}. The brightness $B$ is the phase-space density of the photon flux $F$ (number of photons emitted per second per $0.1\%$ bandwidth), evaluated in the forward direction and at the center of the source \cite{KIM198744}: $B=\frac{d^4 F}{d\theta d\psi dx dz}$, with $\theta$ and $\psi$ the horizontal and vertical angles, $x$ and $z$ the horizontal and vertical coordinates respectively. 
Assuming Gaussian photon distribution, and neglecting the variation of the electron transverse position due to the oscillations through the insertion device, the brightness can be written as \cite{brightnessorigin}:
\begin{equation}
B=\frac{F}{4\pi^{2} \Sigma_{x} \Sigma_{x^{\prime}}\Sigma_{z} \Sigma_{z^{\prime}}} \nonumber
\label{brightness}
\end{equation} 
with $\Sigma_{x, z}$ ($\Sigma_{x^{\prime}, z^{\prime}}$) the size and (the divergence) of the photon beam resulting from the convolution of the electron beam size (divergence)  with the photon emission of a single electron. The beam size and divergence can be expressed as: $\Sigma_{x,z}= \sqrt{\sigma_{x, z (e^{-})}^{2}+\sigma_{x, z (photon)}^{2}}$ and $\Sigma_{x^{\prime}, z^{\prime}}= \sqrt{{\sigma_{x^{\prime}, z^{\prime}(e^{-})}^{2}+\sigma_{x^{\prime}, z^{\prime}(photon)}^{2}}}$, where $\sigma_{x^{\prime},z^{\prime}(photon)}=\sqrt{\frac{\lambda}{2L}}$ the photon beam divergence resulting from single electron emission through an insertion device, $\sigma_{x,z(photon)}=\frac{\sqrt{2\lambda L}}{2\pi}$ the photon beam size \cite{Tanaka:ie5029}, $\lambda$ the wavelength of the emitted radiation and $L$ is the length of the device.

In a non-zero dispersion straight section for a Gaussian electron beam distribution, the total horizontal beam size $\sigma_{x}$ and divergence $\sigma_{x^{\prime}}$ include contributions from both the betatron and the energy spread $\sigma_e$: 
\begin{equation}
\label{eq:last}
\begin{aligned}
\sigma_{x}=\sqrt{\epsilon_{x} \beta_{x} + (\eta_{x}\sigma_{e})^{2}}\\
\sigma_{x^{\prime}}=\sqrt{\frac{\epsilon_{x}}{\beta_{x}} + (\eta^{\prime}_{x}\sigma_{e})^{2}}
\end{aligned}
\end{equation}
 with $\epsilon_{x}$ the horizontal emittance, $\beta_{x}$ the horizontal betatron amplitude function, $\eta_{x}$ and $\eta_{x}^{\prime}$ the horizontal dispersion function and its derivative respectively. In the vertical plane, if the vertical dispersion function $\eta_{z}$ and its derivative $\eta_{z}^{\prime}$ are zero, the vertical beam size $\sigma_{z}$ and divergence $\sigma_{z^{\prime}}$ are expressed as:
 \begin{equation}
 \begin{aligned}
 \sigma_{z}=\sqrt{\epsilon_{z} \beta_{z}}\\
 \sigma_{z^{\prime}}=\sqrt{\frac{\epsilon_{z}}{\beta_{z}}}\\
 \end{aligned}
 \end{equation}with $\epsilon_{z}$ the vertical emittance and $\beta_{z}$ the vertical betatron amplitude function.

At a given wavelength $\lambda$, the emittance of the photon beam $\epsilon_{ph}$ at the diffraction limit is defined by : 
\begin{equation}
\epsilon_{ph}= \sigma_{x,z(photon)}. \sigma_{x^{\prime}z^{\prime}(photon)}=\frac{\lambda}{4\pi} \nonumber
\end{equation}
 
To reach the diffraction limit, the electron beam emittance $\epsilon_{x,z}$ should satisfy the condition: $\epsilon_{x,z} \leq \frac{\lambda}{4\pi}$. For an operation at $\lambda=1$ Angstrom, an electron beam emittance of lower than $10$ pm.rad in both planes is required. Besides, the orientation of the phase space ellipse of the electron beam should match that of the photon beam emitted by a single electron. 

In third generation synchrotron light sources the electron beam emittance (e.g. $3.9$ nm.rad at SOLEIL  \cite{nadji2011operation},  $2.7$ nm.rad at Diamond \cite{diamond}), leads to partial transverse coherence in the X-ray range. 
Recently, diffraction limited storage rings have appeared \cite{borland2013progress, Hettel:xe5005}. 
Indeed, different approaches are implemented to further reduce the emittance of existing light sources.

The natural horizontal emittance $\epsilon_{x0}$ at equilibrium between quantum excitation and radiation damping for an isomagnetic lattice is \cite{lee2004accelerator, WIEDEMANN198824}:
\begin{equation}
\epsilon_{x0}=\frac{1}{J_{x}}\frac{\oint\frac{H_{x}(s)}{\rho}ds}{\oint\frac{1}{\rho^{2}ds}}
\label{eq:emittancedamping}
\end{equation}with $\rho$ is the radius of curvature and $H_{x}(s)$ is the dispersion invariant given by \cite{WIEDEMANN198824}:
\begin{equation}
H_{x}(s)=\gamma_{x} \eta_{x}^{2}(s)+2\alpha_{x} \eta_{x}(s) \eta_{x}^{\prime 2}(s) +\beta_{x} \eta_{x}^{\prime 2}(s) \nonumber
\end{equation} where $\gamma_{x}$, $\beta_{x}$ and $\alpha_{x}$ are Twiss parameters \cite{COURANT19581}. The natural horizontal emittance $\epsilon_{x0}$ for an isomagnetic lattice can be rewritten in the simplified form as \cite{flattice}:
\begin{equation}
\epsilon_{x_{0}}=F(lattice) \frac{E^{2}}{N^{3}}
\label{eq:emittance2}
\end{equation}where $F(lattice)$ is a constant that depends on the lattice design, $N$ is the number of identical dipoles in the storage ring. 

A first approach for reducing the horizontal emittance is given by increasing the number of dipoles in the storage ring, moving from Double Bend Achromat lattice (DBA) \cite{dbachromate} (two dipole magnets with focusing quadrupoles between them to form an achromatic cell) and Triple Bend Achromat (TBA) \cite{dbachromate} (combination of a DBA with a dipole at the center) to Multiple Bend Achromat (MBA) lattice \cite{einfeld1995design}. Among others, there are several examples: 
MAX IV operating with 7-BA lattice and transverse gradient in the dipoles produces a beam with horizontal emittance of $250$ pm.rad \cite{PhysRevSTAB.12.120701, tavares2014max, eriksson2016commissioning}, the project of SPring-8 upgrade, first based on a 10-BA \cite{tsumaki2006very} and later on a 5-BA \cite{tanaka2016spring} to achieve a horizontal emittance of $150$ pm.rad,  the ESRF upgrade proposing a new design based on 7-BA lattice to reach a horizontal emittance of $150$ pm.rad \cite{doi:10.1080/08940886.2014.970931, esrf2}, etc. After the shutdown of the Tevatron collider, it is proposed to use its large tunnel ($6.28$ km circumference) to house a storage ring of a new light source called the Tevatron-Sized Ultimate Storage Ring (TeVUSR) \cite{borland2012exploration}, that thanks to this large circumference, expects a horizontal emittance of $4$ pm.rad using a 7-BA lattice. 

A second approach for horizontal emittance reduction consists in increasing the damping rate by installing damping wigglers in zero dispersion straight sections \cite{WIEDEMANN198824} to enhance the radiation damping by contributing to the synchrotron integrals. One can introduce the ratio $F_{w}=\frac{I_{2w}}{I_{2a}}$ where $I_{2a}$ is the second synchrotron integral without the wiggler and $I_{2w}$ is the additive term due to the wiggler \cite{PhysRevSTAB.4.021001}. The total horizontal emittance with the damping wiggler can be written as:

\begin{equation}
\epsilon_{x}=\epsilon_{xa}\frac{J_{x0}}{J_{x0}+F_{w}}+\epsilon_{xw}\frac{F_{w}}{J_{x0}+F_{w}}
\end{equation}where $\epsilon_{xa}$  and $J_{x0}$ are the emittance and the damping partition number produced in the absence of wiggler, respectively. $\epsilon_{xw}$ is the emittance generated by the wiggler in the limit of $F_{w}\rightarrow \infty$.
The required wiggler length $L_{w}$ for an isomagnetic lattice is given by:
\begin{equation}
\label{eq:dampingwiglength}
L_{w}=\frac{6C (B \rho)^{2}}{r_{e}c \tau_{z}\gamma^{3}B_{w}^{2}}\frac{F_{w}}{1+F_{w}}
\end{equation} with $C$ the machine circumference, $B \rho$ the magnetic rigidity, $r_{e}$ the classical electron radius, $c$ the speed of light  in vacuum, $\tau_{z}$ the damping time, $\gamma$ the relativistic factor, $B_{w}$ the wiggler field.
Different light sources adopted this solution among which PETRA III \cite{tischer2005damping} that reached a horizontal emittance of $1$ nm.rad by installing $80$ m long damping wigglers, PEP-X that replaced the PEP-II tunnel and reached a horizontal emittance of $0.14$ nm.rad by installing $90$ m of damping wigglers \cite{WIEDEMANN198824}, NSLS II at Brookhaven National Laboratory installed $21$ m of damping wigglers in order to achieve $1$ nm.rad horizontal emittance \cite{guo2009nsls}.

The two  strategies, i.e. adding damping wigglers and increasing the number of dipoles to low emittance lattices, can also be combined. In this respect, an "ultimate" PEP-X lattice has been proposed based on 7-BA lattice together with a 90 m damping wigglers in one of the straight sections to reduce the horizontal emittance to $11$ pm.rad \cite{nosochkov2011lattice} at zero current. However, the drawbacks of the damping wigglers solution is the requirement of long insertion devices and the energy consumption of the RF system.

A third approach to reduce the horizontal emittance is given by adjusting the damping partition number by installing in a non-zero dispersion straight section a Robinson wiggler (RW) \cite{robinson}, i.e.  a magnetic system of high field transverse gradient superimposed to its main alternated pole wiggler field. It consists in installing a magnetic system producing a periodic vertical magnetic field $B_{z}$ and transverse field gradient $dB_{z}/dx$. It is enough to install this wiggler in non-zero dispersion straight section without any need to modify the existing infrastructure. First observations have been carried out with success at Cambridge Electron Accelerator \cite{cea}, and in the PS at CERN \cite{BACONNIER1985244} showing $50\%$ horizontal emittance reduction. RW has been recently adopted by other light sources to reduce the horizontal emittance like Heifei Light Source \cite{heifi}, Shanghai
Synchrotron Radiation Facility \cite{tian2017low} and the Metrology Light Source that benefited from RW to improve Touschek lifetime by lengthening the bunch \cite{goetsch2014robinson, MLSPhD}.

In this paper, we show that the Robinson wiggler approach applied to the case of the SOLEIL storage ring enables the reduction of the horizontal emittance, while increasing the energy spread. The experiment has been performed using four U20 undulators as a RW-like by creating off-axis displacement of the electron beam horizontally and simultaneously in the four undulators for getting the required field gradient product. Effects on the radiation are also derived.

\section{Theoretical approach: horizontal emittance reduction by damping partition number adjustment}

\subsection{Transverse and longitudinal properties}

Over one turn in the storage ring, the electron loses energy by emitting radiation, and gains energy from the RF system. Assuming a sinusoidal RF voltage, the electron motion behaves as a damped harmonic oscillator \cite{lee2004accelerator}. The damping process is characterized by the so-called damping partition $D$ that determines the damping rate of the emitted radiation. Considering a storage ring of equilibrium horizontal emittance $\epsilon_{x0}$, the variation of the horizontal emittance $\epsilon_{x}$ due to the modification of the damping partition $D$ at equilibrium between quantum excitation and radiation damping can be expressed as  \cite{abualrob2012soleil}:

\begin{equation}
\epsilon_{x}=\epsilon_{x0}\frac{1}{1-D}
\label{eq:emittance}
\end{equation}

The damping partition numbers $J_{x}$, $J_{z}$, $J_{s}$, characterizing the damping rate, are defined as \cite{lee2004accelerator}:
\begin{equation}
J_{x}=1-D, \hspace{5 mm} J_{z}=1, \hspace{5 mm} J_{s}=2+D \nonumber
\end{equation}

According to Robinson theorem \cite{robinson}, the total sum of damping partition numbers i.e. of the amount of damping decrement over all degrees of freedom is constant: $\sum J_{i}=J_{x}+J_{z}+J_{s}=4$ .
In terms of the synchrotron integrals $I_{2}$ and $I_{4}$, $J_{x}$ is expressed as $J_{x}=1-(I_{4})/(I_{2})$  \cite{lee2004accelerator}, where $I_{2}$ and $I_{4}$ are given by:
\begin{equation}
\begin{cases}
I_{2}=\oint \frac{1}{\rho^{2}}ds\\  \nonumber
I_{4}=\oint \frac{\eta_{x}(s)}{\rho} \left( \frac{1}{\rho^{2}}-2K(s) \right) ds\\
\end{cases}
\end{equation}where $K(s)$ is the normalized focusing strength in the dipole. It is given in terms of the dipole field $B_{z}$ and its transverse variation $dB_{z}/dx$ (usually known as the field gradient) as $K(s)=(1)/(B_{z}\rho)(dB_{z}/dx)$.

At equilibrium between quantum excitation and radiation damping, the relative energy spread for an isomagnetic lattice is \cite{lee2004accelerator}:
\begin{equation}
\sigma_{e}^{2}=\frac{C_{q}\gamma^{2}}{J_{s}\rho}
\end{equation}

The variation of the energy spread $\sigma_{e}$ as a function of $D$ can be written as follows \cite{abualrob2012soleil}:

\begin{equation}
\sigma_{e}^{2}=\sigma_{e0}^{2}\frac{2}{2+D}
\label{eq:espread}
\end{equation}  

According to equations \ref{eq:emittance} and \ref{eq:espread}, the horizontal emittance $\epsilon_{x}$ and the relative energy spread $\sigma_{e}$ can be modified by varying the horizontal and the longitudinal damping partition numbers $J_{x}$ and $J_{s}$ respectively; i.e. by varying the damping partition $D$. The damping partition for a given lattice is determined by its design. It is given in terms of the magnetic field  $B_{z}$ and the field gradient $dB_{z}/dx$ by:
\begin{equation}
D=\frac{\frac{1}{2\pi} \left( \oint \frac{\eta_{x}}{\rho^{3}}ds + \frac{2}{B^{2}\rho^{2}} \oint \eta_{x} B_{z} \frac{dB_{z}}{dx} ds \right) }{\oint \frac{ds}{\rho^{2}}}
\label{eq:generaldampingpartition}
\end{equation}

For an isomagnetic storage ring with identical bending magnets, and equipped with separate function magnets (i.e. bending magnets for deflection and quadrupoles for focusing) such as SOLEIL, the damping partition is given by: 
\begin{equation}
D=\frac{\alpha R}{\rho}  \nonumber
\end{equation} where $R$ is the radius of the storage ring and $\alpha$ is the momentum compaction factor. For the case of SOLEIL ($\alpha=4.16 \times 10^{-4}$, $R=57.37$ m, $\rho=5.3$ m), the damping partition $D \approx 0$.

If $D$ can be reduced from its usual value $D \approx 0$ to $D=-1$, the horizontal emittance can be divided by $2$ (see equation \ref{eq:emittance}), while the energy spread will be increased by $\sqrt{2}$ (see equation \ref{eq:espread}).

The bunch length represents the longitudinal distribution of the electron in the phase space and is related to energy spread. In the zero current regime, the bunch length $\sigma_{t}$ is given by \cite{bunchsands}:
\begin {equation}
\sigma_{t}=\frac{\alpha}{2\pi f_{s}}\sigma_{e}
\label{eq:bunchlength}
\end{equation} where $f_{s}$ is the synchrotron frequency expressed in terms of the radio-frequency voltage $V$, the momentum compaction factor $\alpha$, the harmonic number $h$, the electron energy $E$, and the revolution frequency $f_{0}$ as:

\begin{equation}
f_{s}=f_{0} \sqrt{\frac{V \alpha h}{2 \pi E}}
\end{equation}

\subsection{Requirements in terms of magnetic field}

For an isomagnetic lattice, the damping partition given by equation \ref{eq:generaldampingpartition} can be reduced to:
\begin{equation}
D=\frac{\rho \langle {\eta_{x}}\rangle_{s}}{\pi (B \rho)^{2}} \int_0 ^ {L_{w}} {B_{z}\frac{dB_{z}}{dx}ds}
\label{eq:simpledampingpartition}
\end{equation}

The damping partition $D$ can be significantly modified by installing an insertion device of high vertical field and high field gradient in a non-zero dispersion straight section. A RW of length $L_{w}$ and peak field $B_{w}$ contributes to the modification of $D$ if inserted in a straight section whose average dispersion function over the length of the wiggler is $\langle \eta_{x} \rangle$. The damping partition can get a negative value if $B_{w}\frac{dB_{w}}{dx}<0$ according to equation \ref{eq:simpledampingpartition}.

 The SOLEIL storage ring hosts three types of straight sections, as presented in Table \ref{table:ss}, all providing non-zero dispersion. $D$ could be reduced from $\approx 0$ to $-1$ by installing a wiggler of $\int B_{w} \hspace{1 mm} (dB_{w}/dx) \hspace{1 mm} ds = 193.4$ $T^{2}$ in a short straight section ($\eta_{x}=0.252$ m), leading to a reduction of the horizontal emittance from its present value of $ 3.9$ nm.rad to $1.95$ nm.rad. Inversely, the energy spread will be increased from the present value of $1.01 \times10^{-3}$ to $1.43 \times10^{-3}$.

\begin{table}
\caption{\label{table:ss} Length of straight sections at SOLEIL and optical functions at the center} 
\begin{ruledtabular}
\begin{tabular}{ccccc}
Straight section   & Length (m) & $\eta_{x}$ (m) & $\beta_{x}$ (m)& $\beta_{z}$ (m)\\ 
\hline
Long         & 4$\times$12            & 0.206 &   5.58  &  8.03       \\
Medium    & 12$\times$7          &0.165  &   4.6       &  2.24      \\
Short     & 8$\times$3.6            & 0.252 &   14.38  &  2.36      \\
\end{tabular}
\end{ruledtabular}
\end{table}

\section{Experimental observation of the Robinson effect at SOLEIL}
 
\begin{figure} 
\includegraphics{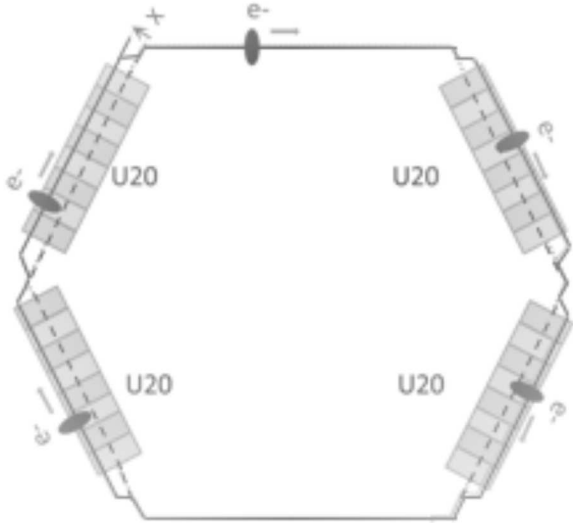}
\caption{\label{Fig:bump}
Schematic presentation of the simultaneous horizontal bumps in the four U20 undulators. Dashed line: the on-axis electron beam trajectory, continuous line: the displaced trajectory by applying a horizontal bump of value x.}  
\end{figure}

To bring out the effect of a RW on the emittance and the relative energy spread at SOLEIL, different experiments have been performed using four already existing in-vacuum undulators (four U20s) \cite{benabderrahmane2007commissioning, couprie2010panoply} in the storage ring. The four U20s are all installed in dispersive short straight sections and generate periodic vertical magnetic field and significant off-axis field gradient. To maximize the total effect, the four U20s were used simultaneously. Figure \ref{Fig:bump} shows the schematic experimental setup. The electron beam is displaced off-axis in the four undulators by applying simultaneous horizontal bumps using horizontal dipolar correctors at the entrance and at the exit of each undulator. At the entrance the electron beam is displaced horizontally parallel to its initial trajectory and comes back to the initial one at the exit of the undulator. After each horizontal simultaneous bump, the tunes are set back to the nominal values using two quadrupole families. Horizontal beam size and bunch length are measured at each bump with the four U20s open at maximum gap (bare machine) and with the four undulators closed at minimum gap of 5.5 mm.

To study the Robinson effect, the horizontal beam size and the bunch length are measured at 18 mA current distributed in 416 bunches. The low current per bunch enables operation of the machine close to the zero current regime so as to satisfy equation \ref{eq:bunchlength}. 


\subsection{The U20: a RW-like undulator}

\begin{figure} 
\includegraphics{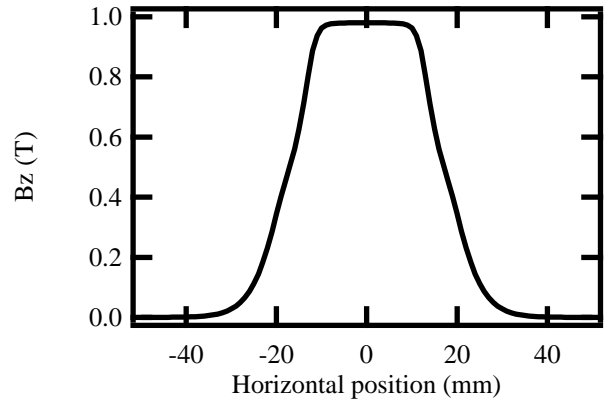}
\caption{\label{Fig:BU20}
Peak magnetic field of a U20 undulator calculated as a function of the horizontal beam position with RADIA for the parameters listed in Table \ref{table:U20robinson} (minimum gap).}  
\end{figure}

Figure \ref{Fig:BU20} illustrates the peak field variation over the horizontal range $\pm 50$ mm for a U20 undulator calculated with RADIA code \cite{elleaume1997computing, Chubar:hi3178} for the parameters given in Table \ref{table:U20robinson}. 
The magnetic field of a U20 is constant over $\pm 10$ mm in the vicinity of the central on-axis position. Beyond $\pm10$ mm, the peak field drops sharply creating a strong magnetic field gradient. The superimposition of the strong magnetic field to the strong field gradient as in a RW can be achieved by creating off-axis propagation of the electron beam through the U20. Consequently, a U20 undulator is a good candidate to study Robinson effect thanks to the particular transverse variation of its magnetic field over wide horizontal range. 

\begin{table}
\caption{\label{table:U20robinson} Main characteristics of the in-vacuum U20 undulators used for the experimental study of the Robinson effect}
\begin{ruledtabular}
\begin{tabular}{ccc}
Characteristic & Value & Unit \\ 
\hline
Type &in-vacuum & -\\
Magnet material&$Nd_{2}Fe_{17}B$&-\\
Magnet dimensions (s, x, z)&7.5, 50, 30& mm\\
Magnet chamfer size& 4$\times$4& mm\\
Magnetization $B_{r}$& 1.05 & T\\
Pole material&Vanadium Permendur&-\\
Pole dimensions (s, x, z)&2.5, 33, 22& mm\\
Pole saturation field $B_{s}$&2.35&T\\
Pole chamfer size& 4$\times$4&mm\\
Peak field & 0.97 & T   \\
Period length    &    20 & mm   \\
(Magnetic) minimum gap height &  5.5 &  mm   \\
Period number & 98 & -\\
Deflection parameter  & 1.8 & -\\
\end{tabular}
\end{ruledtabular}
\end{table}

\begin{figure} 
\includegraphics{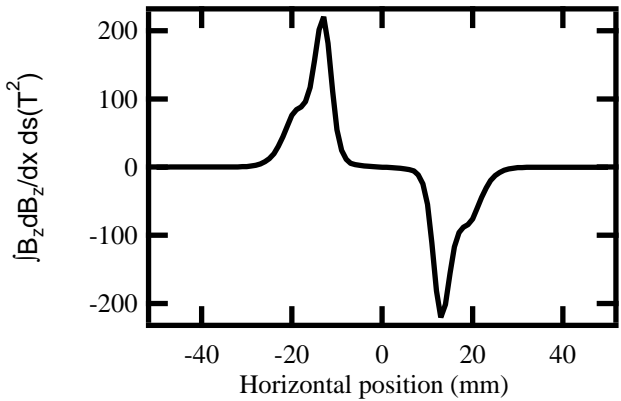}
\caption{\label{Fig:BdBdx}
Transverse variation of $\int{B_{z}(dB_{z}/dx) ds}$ calculated with RADIA for the four U20 undulators closed at gap 5.5 mm.}  
\end{figure}

 Figure \ref{Fig:BdBdx} illustrates the magnetic field multiplied by the field gradient integrated over the undulator length $\int{B_{z}(dB_{z}/dx) ds}$ for the four U20 undulators versus transverse position. The zone of interest for the observation of the Robinson effect is that of maximum magnetic field and maximum field gradient. The four U20 undulators show a significant $\int{B_{z}(dB_{z}/dx) ds}$ peak of $\pm 220$  $T^{2}$ at $x=\pm13$ mm.

\subsection{Machine tuning for off-axis propagation}

\begin{figure}
\includegraphics{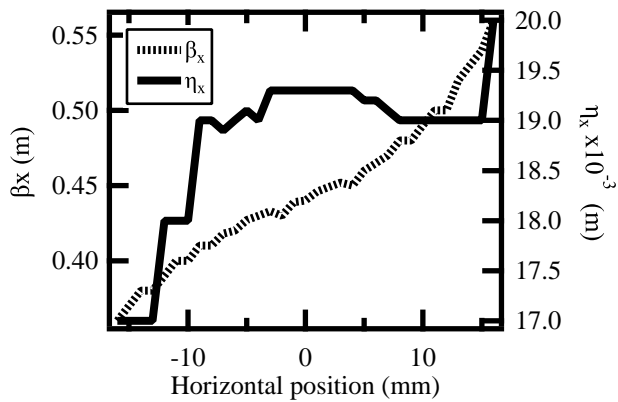}
\caption{\label{Fig:betaeta}Variation of the horizontal betatron and dispersion functions with the horizontal position in the four U20 undulators at the pinhole camera location. Functions obtained with AT simulation code.}
\end{figure}

In the case of the SOLEIL storage ring, the optical functions vary significantly when horizontal bumps reach large values, since the horizontal bump creates off-axis propagation in a large number of sextupoles leading to additional focusing. The optical functions corresponding to each bump are simulated with the Accelerator Toolbox (AT) code \cite{terebilo2001accelerator} at the location of the photon beam profile sensor as a function of the horizontal beam position at any given horizontal position in the four U20s and any longitudinal position in the storage ring. . Figure \ref{Fig:betaeta} shows the AT simulation of the optical functions $\eta_{x}(x)$ and $\beta_{x}(x)$, assuming a simultaneous horizontal bump in the four U20 undulators. 

\begin{figure} 
\includegraphics{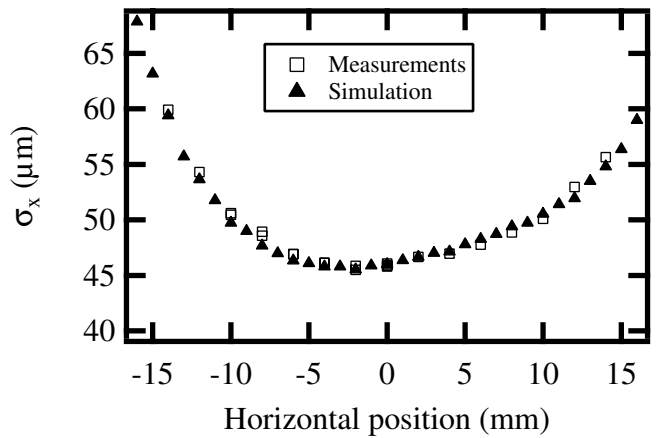}
\caption{\label{Fig:verification}
Horizontal beam size variation for a bare machine as a function of the horizontal position in the four U20 undulators. $\blacktriangle$: orbit bump simulation with AT code, $\square$: measurements with the pinhole camera with a 5 $\mu m$ precision.}  
\end{figure}

\begin{figure} 
\includegraphics{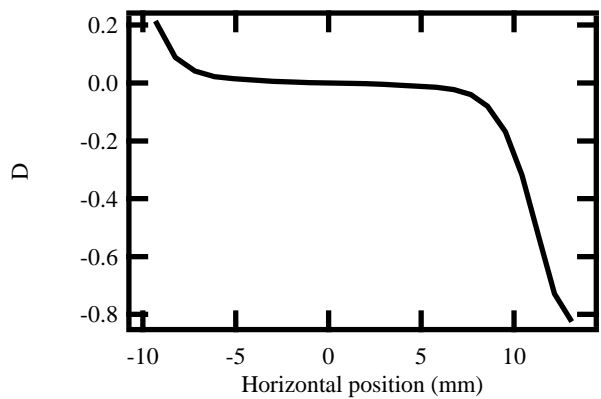}
\caption{\label{Fig:partitionumber}
Damping partition calculated with equation \ref{eq:simpledampingpartition} as a function of the horizontal position when the four U20 undulators are closed at minimum gap. $\langle\eta_{x}\rangle$: the average horizontal dispersion function obtained by AT simulation. $B \rho$: the magnetic rigidity $B\rho=9.13$ T.m. $\int_{0} ^ {L} {B_{z} (dB_{z}(x)/dx) \hspace{2 mm}ds}$: The integral of the magnetic field times the field gradient for the four U20 undulators calculated with RADIA code (see Fig. \ref{Fig:BdBdx}).}  
\end{figure}

\begin{figure*}
\includegraphics{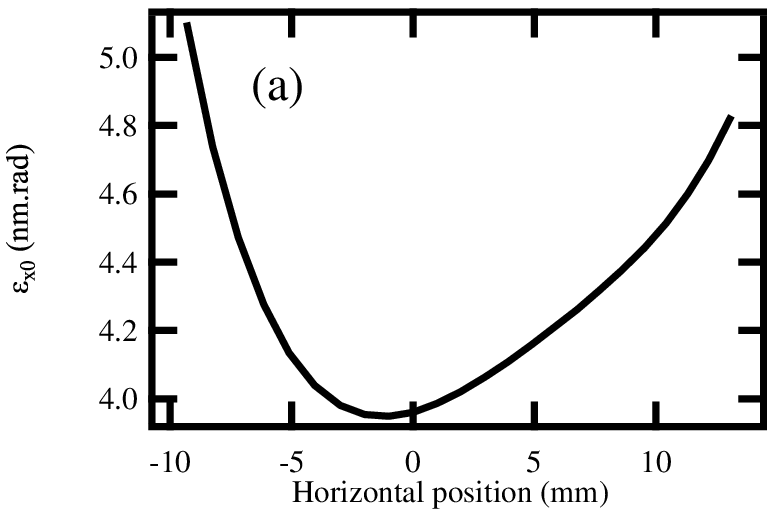}
\quad
\includegraphics{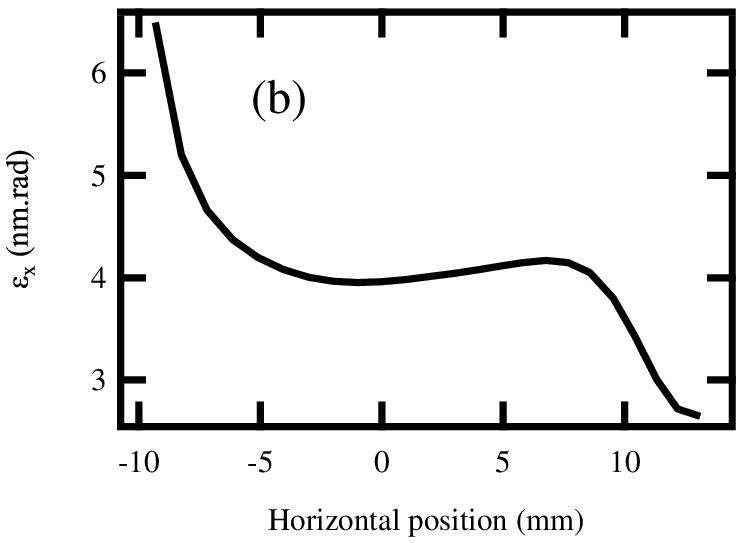}
\caption{\label{Fig:epsilon} (a): Natural horizontal emittance variation as a function of the horizontal beam position in the four U20 undulators obtained with AT simulation (U20s are open at maximum gap). (b): calculated horizontal emittance variation as a function of the horizontal position in the four U20 undulators (U20s are closed at minimum gap). Calculations performed with equation \ref{eq:emittance} with $\epsilon_{x0}(x)$: natural horizontal emittance corresponding to the horizontal beam position according to AT simulation and $D(x)$: the damping partition (see Fig. \ref{Fig:partitionumber}).}
\end{figure*}

Figure \ref{Fig:verification} shows the horizontal beam size measured with a pinhole camera as a function of the horizontal position for a bare machine and the beam size obtained by simulation. Good agreement between measurements and AT results confirms the good quality of the model used in AT code.

Figure \ref{Fig:partitionumber} shows the damping partition $D$ as a function of the horizontal position in the four U20s (undulators are closed at gap $5.5$ mm) calculated with equation \ref{eq:simpledampingpartition}. The dissymetry is due to the fact that $\langle \eta_{x}(x) \rangle$ differs for positive and negative values of the horizontal position. The damping partition can even evaluate to negative values and reaches a minimum of $-0.8$, whereas a damping partition of $-1$ is required to reduce the horizontal emittance by factor of 2 (see equation \ref{eq:emittance}). It is not possible to get the value of $D=-1$, since some beam instability appears at larger bumps leading to horizontal emittance reduction by a factor of 0.45 and energy spread increase by a factor of 1.3.

The variation of the natural horizontal emittance with the bump is also taken into account. Figure \ref{Fig:epsilon} (a) shows the natural horizontal beam emittance as a function of the horizontal beam position obtained with AT simulation (U20s are open). It is modified due to the horizontal beam displacement, asymmetric increase is noticed around the nominal propagation axis. Figure \ref{Fig:epsilon} (b) shows the calculated horizontal beam emittance with equation \ref{eq:emittance} in the case of four U20s closed at minimum gap, with the natural horizontal beam emittance $\epsilon_{x0}(x)$ deduced from Fig. \ref{Fig:epsilon} (a) and the damping partition $D(x)$ of Fig. \ref{Fig:partitionumber}. 

\subsection{Measurement of Robinson effect on the longitudinal beam properties}
\label{sec:longprop}

\begin{figure} 
\includegraphics{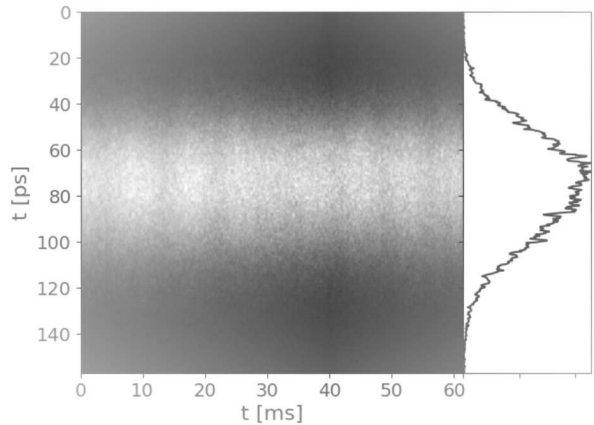}
\caption{\label{Fig:bunchphoto}
Time evolution of the longitudinal bunch profile detected by streak camera. Single bunch mode, $I=0.7$ mA, $V_{RF}=1.37$ MV and electron beam positioned at $x=16$ mm in the four U20s.}  
\end{figure}

\begin{figure}
\includegraphics{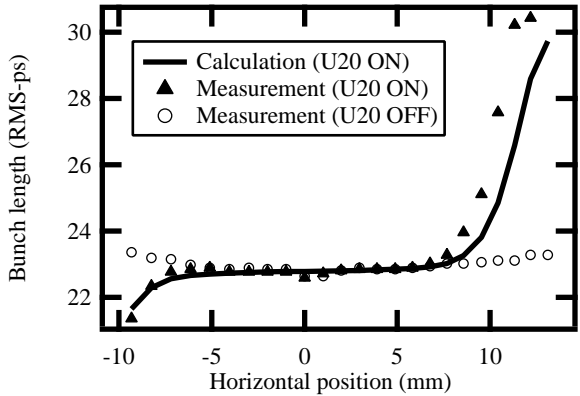}
\caption{\label{Fig:bunchcalmes}Electron bunch length variation as a function of the horizontal position in the four U20 undulators. U20 ON: U20 undulators set at minimum gap ($5.5$ mm), U20 OFF: U20 undulators set at maximum gap ($30$ mm). Calculations: \protect\rule{1cm}{2pt}, from measurements with the streak camera for the U20s set at minimum gap: $\blacktriangle$, at maximum gap: $\circ$. Streak camera of 2 ps FWHM resolution.}
\end{figure}

The bunch length is measured as a function of the horizontal position in the four U20s with a HAMAMATSU-C10910 Series streak camera \cite {tordeux2007ultimate} of 2 ps-FWHM resolution to infer the effect on the energy spread (equation \ref{eq:bunchlength}). Figure \ref{Fig:bunchphoto} shows a typical bunch image detected by the streak camera. Figure \ref{Fig:bunchcalmes} compares the measured bunch length with that obtained with equation \ref{eq:bunchlength} using the damping partition due to the four U20s closed at minimum gap found in Fig. \ref{Fig:partitionumber}. Bunch length calculations assumes $\alpha=416.4 \times 10^{-6}$ to be constant as a function of the horizontal position as predicted by AT simulation, $V=1.2$ MV, $h=416$, $E=2.75$ GeV, $f_{0}=847$ kHz and $\sigma_{e}$ the energy spread calculated with equation \ref{eq:espread} assuming constant natural energy spread ($\sigma_{e0}=1.01$ $\times$ $10^{-3}$) over the whole horizontal range as predicted by AT simulation, and the damping partition found in Fig. \ref{Fig:partitionumber}.

Measurements confirm the expected increase of bunch length when U20s are closed at minimum gap due to the increase of the energy spread. However, measurements are less in accordance with calculations at large bump values because during the experiment it was not possible to displace the electron beam so that its horizontal position is exactly the same in the four U20s. 

\subsection{Measurement of Robinson effect on the transverse beam properties}

\begin{figure}
\includegraphics{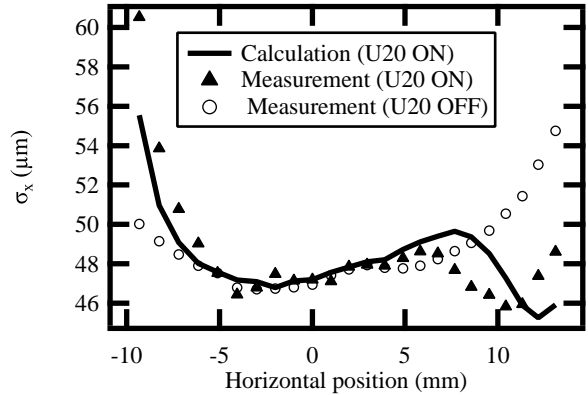}
\caption{\label{Fig:sigmacalmes}Calculated and measured horizontal beam size variation at the pinhole camera as a function of the horizontal position in the four U20 undulators. U20 ON: U20 undulators set at minimum gap ($5.5$ mm), U20 OFF: U20 undulators set at maximum gap ($30$ mm). Calculations: \protect\rule{1cm}{2pt},  measurements with the U20s set at minimum gap: $\blacktriangle$, at maximum gap: $\circ$.}
\end{figure}

Figure \ref{Fig:sigmacalmes} shows the horizontal beam size as a function of the horizontal beam position in the four U20s measured with a pinhole camera \cite{labat2007streak} of 5 $\mu m$ precision to deduce the effect on the horizontal beam emittance (equation \ref{eq:last}), taking into account the corresponding measured energy spread and the horizontal beam size expected from theoretical calculations using equation \ref{eq:last}, where $\sigma_{e}$ is the calculated energy spread (see Fig. \ref{Fig:bunchcalmes}), $\eta_{x}$ and $\beta_{x}$ are the optical functions at the pinhole camera obtained with AT simulation (see Fig. \ref{Fig:betaeta}) and $\epsilon_{x}$ is the horizontal beam emittance calculated with equations \ref{eq:emittance} and \ref{eq:simpledampingpartition} and the optical functions $\eta_{x}$ and $\beta_{x}$ (see Fig. \ref{Fig:epsilon}). 

The horizontal beam size reduction due to the Robinson effect expected at $x=12$ mm is confirmed by experimental observations. At large bump values, a slight shift between measurements and calculations appears because the horizontal beam position is not exactly the same in the four U20s.

\section{Expected spectral performance}

Following the experimental observation of Robinson effect, the radiation properties of already installed undulators are studied under the modifications introduced by a RW with SRW (Synchrotron Radiation Workshop) code \cite{chubar1998accurate}. The undulator HU640 operating at low photon energy range in the linear horizontal polarization, and the U20 undulator operating at high photon energy range, are supposed to be installed in a new machine whose emittance and energy spread are modified by a RW while assuming unperturbed dispersion and betatron functions. (see Table \ref{tab:HU640U20} for the main characteristics of both undulators).

\begin{table}[ht]
\caption{\label{tab:HU640U20}Characteristics of the HU640 and U20 undulators}
\begin{ruledtabular}
\begin{tabular}{ccc}
Undulator &     HU640 & U20\\
\hline
Technology &          Electromagnetic &  HPM \footnote{HPM: Hybrid Permanent Magnet}, in-vacuum \\
$(B_{z})_{max}$ & 0.15 T &1.08 T \\
Period length  & 640 mm &   20 mm\\
$K$  & 8.95      &  2\\
Energy range  & (5-40) eV & (3-18) keV\\
Straight section   & Long     & Short \\
\end{tabular}
\end{ruledtabular}
\end{table}

\begin{figure*}[ht!]
\includegraphics{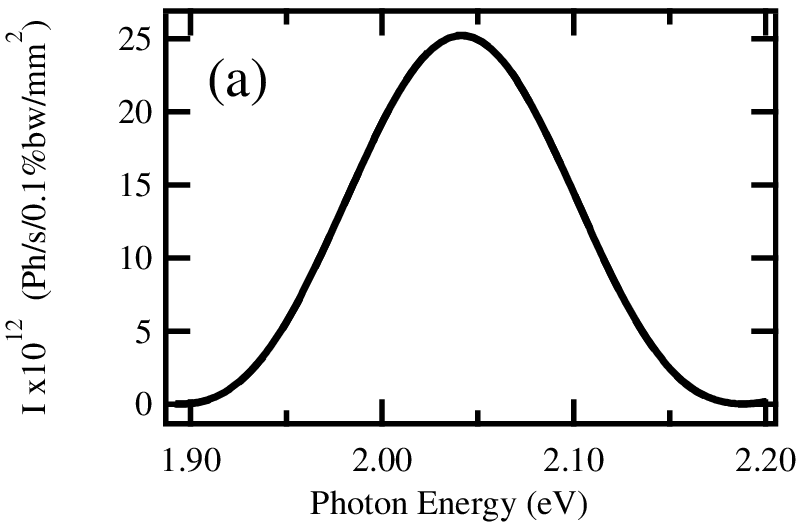}%
 \quad
\includegraphics{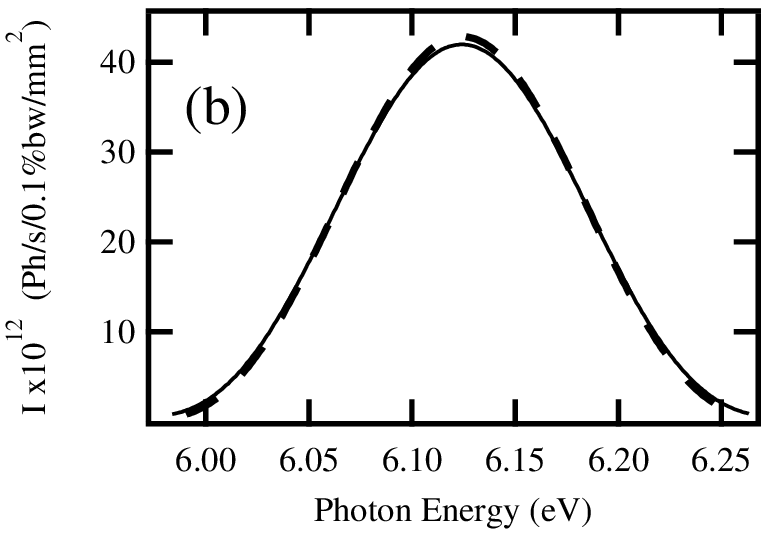}
 \caption{\label{Fig:HU640}%
Flux density emitted by the undulator HU640 ($K=8.95$) calculated with SRW through an aperture of 0.1 $\times$ 0.1 mm$^{2}$ located at $10$ m from the source. (a): zoom on the harmonic H1, (b) zoom on the harmonic H3. \protect\rule{1cm}{2pt} $\epsilon_{x}=1.95$ nm.rad, $\sigma_{e}=1.428$$\times$$10^{-3}$. \protect\rule{0.2cm}{2pt} \hspace{0.001cm} \protect\rule{0.2cm}{2pt} $\epsilon_{x}=3.9$ nm.rad, $\sigma_{e}=1.01$$\times$$10^{-3}$.  Beam energy of 2.75 GeV and beam current of 500 mA.
 }%
\end{figure*}

The effect on the spectrum can be understood by considering the various contributions of the spectral broadening $\left(\frac{\Delta \lambda}{\lambda}\right)_{tot}$. For a mono-energetic filament electron beam, the undulator line presents a natural linewidth $\left(\frac{\Delta \lambda}{\lambda}\right)_{hom}=\frac{0.9}{nN_{0}}$ with $n$ the harmonic number and $N_{0}$ the number of the undulator periods, the inhomogeneous broadening due to the emittance $\left(\frac{\Delta \lambda}{\lambda}\right)_{\epsilon}=\left(\frac{\gamma^2}{1+\frac{K^2}{2}}\right)(\sigma_{x^{\prime}}^{2}+\sigma_{z^{\prime}}^{2})$ with $\gamma$ the relativistic factor and $K$ the deflection parameter of the undulator, the inhomogeneous broadening due to the energy spread $\left(\frac{\Delta \lambda}{\lambda}\right)_{\sigma_{e}}=2\sigma_{e}$ and the inhomogeneous broadening due to the beam size $\left(\frac{\Delta \lambda}{\lambda}\right)_{\sigma}=\left(\frac{\gamma^{2}}{1+\frac{K^{2}}{2}}\right) \left((\frac{\sigma_{x}}{d})^{2} +(\frac{\sigma_{z}}{d})^{2}\right)$ at a distance $d$ where the radiation is collected. Consequently, the total spectral broadening can be expressed as:

\begin{widetext}
\begin{equation}
\left(\frac{\Delta \lambda}{\lambda}\right)_{tot}=\sqrt{\left(\frac{0.9}{nN_{0}}\right)^{2}+\left(\frac{\gamma^2}{1+\frac{K^2}{2}}\right)^{2}(\sigma_{x^{\prime}}^{4}+\sigma_{z^{\prime}}^{4})+\left(2\sigma_{e}\right)^{2}+\left(\frac{\gamma^{2}}{1+\frac{K^{2}}{2}}\right)^{2} \left((\frac{\sigma_{x}}{d})^{4} +(\frac{\sigma_{z}}{d})^{4}\right)}\\
\label{eq:brodening}
\end{equation}
\end{widetext}

Figure \ref{Fig:HU640} compares the calculated flux density emitted by the HU640 undulator considering the present SOLEIL horizontal emittance and energy spread, and that modified due to the assumed presence of a RW. The flux density emitted by the HU640 for the first (H1) and the third (H3) harmonics remains practically unchanged. The harmonic H1 is unaffected whereas the H3 flux is reduced by $0.5\%$. The harmonic widths are mostly determined by the energy spread, since $\left(\frac{\Delta \lambda}{\lambda}\right)_{\sigma_{e}} \approx 2.8 \times 10^{-3}$, while $\left(\frac{\Delta \lambda}{\lambda}\right)_{\epsilon}=4.95 \times 10^{-4}$ and $\left(\frac{\Delta \lambda}{\lambda}\right)_{\sigma}=4.7 \times 10^{-4}$ (calculated at the source point: $\sigma_{x}=257.6$ $\mu$m, $\sigma_{z}=17.7$ $\mu$m, $\sigma_{x^{\prime}}=26.4$ $\mu$rad, $\sigma_{z^{\prime}}=2.2$ $\mu$rad). As the contribution of the energy spread is independent of the harmonic number, the effect of the energy spread on the lower order harmonics combines with the homogeneous linewidth, whereas for higher order harmonics the broadening is mainly determined by the energy spread. Given that the HU640 is composed of 14 periods, then $\left(\frac{\Delta \lambda}{\lambda}\right)_{hom}=7\%$ for H1, and $\left(\frac{\Delta \lambda}{\lambda}\right)_{hom}=2\%$ for H3.

\begin{figure*}
\includegraphics{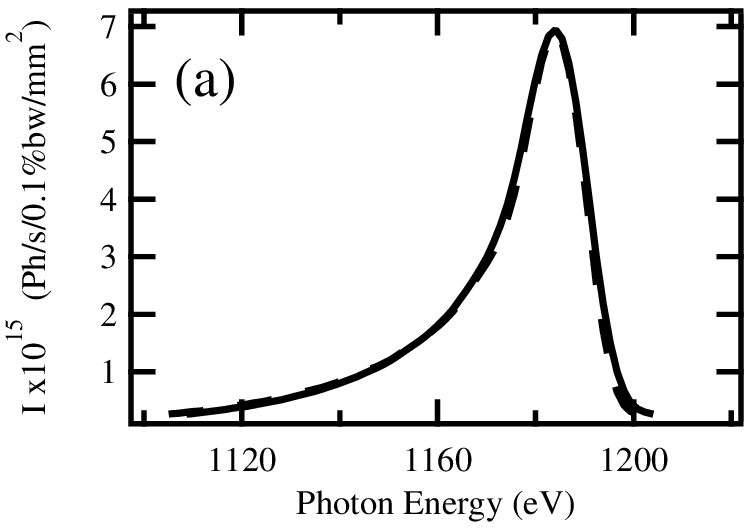}%
\includegraphics{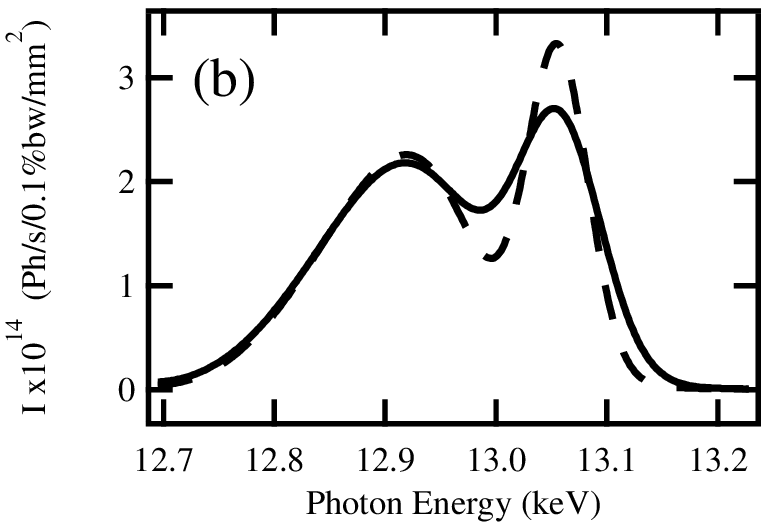}
 \caption{\label{Fig:U20}%
 Flux density emitted by the undulator U20 ($K=2$) calculated with SRW through an aperture of 0.1 $\times$ 0.1 mm$^{2}$ located at $10$ m from the source. (a): zoom on the harmonic H1, (b) zoom on the harmonic H11. \protect\rule{1cm}{2pt} $\epsilon_{x}=1.95$ nm.rad, $\sigma_{e}=1.428$$\times$$10^{-3}$. \protect\rule{0.2cm}{2pt} \hspace{0.001cm} \protect\rule{0.2cm}{2pt} $\epsilon_{x}=3.9$ nm.rad, $\sigma_{e}=1.01$$\times$$10^{-3}$. The beam energy is 2.75 GeV, the beam current is 500 mA.
 }%
\end{figure*}

Figure \ref{Fig:U20} compares the calculated flux density emitted by the U20 undulator operating in the high energy range considering the present SOLEIL horizontal emittance and energy spread and that modified due to the assumed presence of a RW. The flux density of the harmonic H1 is not affected and that of the harmonic H11 is reduced by about $18\%$. The different contributions to the total harmonic broadening are: $\left(\frac{\Delta \lambda}{\lambda}\right)_{\sigma_{e}} \approx 2.8 \times 10^{-3}$, $\left(\frac{\Delta \lambda}{\lambda}\right)_{\epsilon}=9.14 \times 10^{-3}$ and $\left(\frac{\Delta \lambda}{\lambda}\right)_{\sigma}=6.79 \times 10^{-3}$ (calculated at the source point: $\sigma_{x}=265.1$ $\mu$m, $\sigma_{z}=9.5$ $\mu$m, $\sigma_{x^{\prime}}=30.5$ $\mu$rad, $\sigma_{z^{\prime}}=4.1$ $\mu$rad). Concerning the harmonic H1, the homogeneous linewidth dominates the effects of the energy spread ($\left(\frac{\Delta \lambda}{\lambda}\right)_{hom}=10^{-2}$, the U20 is composed of 98 periods). Flux reduction noticed of the harmonic H11 is due to the combinaion between the energy spread and the homogeneous linewidth ($\left(\frac{\Delta \lambda}{\lambda}\right)_{hom}=9 \times 10^{-3}$).


Increasing the energy spread has a greater effect on the harmonic intensity for the U20 undulator compared to the HU640 one due to larger contribution of the emittance and the energy spread to the inhomogeneous broadening in the case of the U20. In addition, the homogeneous broadening is larger for the HU640 than for the U20.

\section{Conclusion}
The Robinson effect was observed and validated experimentally at SOLEIL. This novel development is a critical step forward achieved even without constructing a RW. The experiment was performed by making use of the high field and field gradient of already installed undulators in the storage. The expected Robinson effects on reducing the horizontal emittance and increasing the energy spread are observed in agreement with theoretical expectations: horizontal emittance is reduced by ratio of 35$\%$ and the energy spread is increased by ratio of 30$\%$ with respect to the present values. RW has an impact on the spectral distribution of the photon flux density emitted by insertion devices. In the low energy range it leads to a very tiny photon flux reduction, while in the high energy range it has a much larger effect on the photon flux. This makes a RW not a good candidate for synchrotron radiation facilities, but an excellent candidate for machines providing collision experiments which require beams of tiny dimensions.

\begin{acknowledgments}
The author would like to acknowledge Ryutaro Nagaoka from Synchrotron SOLEIL for reviewing this manuscript and for giving his time for continuous and constructive discussions.
\end{acknowledgments}
 
\nocite{*}
\newpage
\bibliography{apssamp}

\end{document}